\documentstyle[twocolumn,titlepage,openbib]{article}
\begin{document}
\newcommand{\N}{$^{1,2}$}
\newcommand{\sg}{$\sigma$}
\newcommand{\bt}{$\beta$}
\newcommand{\rs}{$\rho_{\sigma}$}
\newcommand{\ri}{$\rho_{0.25}$}
\newcommand{\rg}{$\rho_{0.5}$}
\newcommand{\rc}{$\rho_{1.0}$}
\newcommand{\rd}{$\rho_{2.0}$}
\newcommand{\rp}{$r_P$}
\newcommand{\ras}{$r_S$}
\newcommand{\rk}{$r_K$}
\newcommand{\lf}{Log$f_1$}
\renewcommand{\le}{Log$f_2$}
\newcommand{\pu}{GM$^3$}
\newcommand{\pp}{BG$^2$M$^2$}
\newcommand{\nov}{$>99.99\%$}
\newcommand{\nsg}{$<90\%$}
\newcommand{\mb}{$M_B$}
\newcommand{\be}{\begin{equation}}
\newcommand{\ee}{\end{equation}}

\begin{titlepage}
\begin{center}
\Huge
{\bf
The Local Galaxy Density \\ and the Arm Class \\ of Spiral Galaxies
}

\large

\vspace{4cm}

Giuliano Giuricin\N, Pierluigi Monaco\N,

Fabio Mardirossian\N, Marino Mezzetti\N
\end{center}

\large
\bigskip
\bigskip
\bigskip
\bigskip

(1) Scuola Internazionale Superiore di Studi Avanzati (SISSA), via \linebreak
Beirut~4, 34013 -- Trieste, Italy

(2) Dipartimento di Astronomia, Universit\`a degli studi di Trieste, via
Tiepolo 11, 34131 -- Trieste, Italy

\bigskip
\bigskip

email: \\
giuricin@tsmi19.sissa.it\\
monaco@tsmi19.sissa.it\\
mardirossian@tsmi19.sissa.it\\
mezzetti@tsmi19.sissa.it

\bigskip
\bigskip

\begin{center} SISSA Ref. 102/93/A \end{center}

\end{titlepage}

\begin{abstract}
\large

We have examined the effect of the environmental density on  the arm
classification of an extensive sample of spiral galaxies included in the Nearby
Galaxy Catalog (Tully, 1988a). We have also explored the dependence of the
arm class of a galaxy on other factors, such as its blue absolute magnitude and
its disk-to-total mass ratio, inferred in the literature either from the
gradient  of a good galaxy rotation curve or from a photometric mass
decomposition method.

We have found that the arm class is strongly related to the  absolute magnitude
in the mid-type spirals (in the sense that  grand design galaxies are, on
average, more luminous than  flocculent objects), whilst this relation is
considerably weaker in the  early and late types. In general the influence of
the local density  on the arm structure appears to be  much weaker than that of
the  absolute magnitude. The local density acts essentially in strengthening
the arm class--absolute magnitude relation for the  mid types, whereas no
environmental density effects are observed in  the early and late types.

Using the most recent estimates of the disk-to-total mass ratio, we do not
confirm this ratio to be a significant factor which affects the arm  class;
nevertheless, owing to poor statistics and large uncertanties, the issue
remains open. Neither a local density effect nor an
unambiguous bar effect on the  disk-to-total mass ratio is detectable; the
latter finding   may challenge some theoretical viewpoints  on the formation of
bar structures.

\bigskip
\bigskip

{\it Subject headings:} galaxies: general - galaxies: structure -  galaxies:
internal motions - galaxies: clustering

\end{abstract}

\normalsize
\section{Introduction}

The degree of symmetry and continuity of spiral arms in galaxies is the basis
of the classification system  introduced by  Elmegreen \& Elmegreen (1982).
They assigned galaxies to 12 distinct arm  classes (AC) ranging from AC=1
(fragmented arms with no symmetry) to  AC=12 (two long, sharply defined arms
which dominate the appearance of  the galaxy). A few years later the same
authors published a catalog  of spiral arm classes of 765 galaxies, in which
the original classification  was slightly refined (Elmegreen \& Elmegreen,
1987).

Several properties have been examined for correlations with AC. Grand design
(hereafter G) spirals (i.e. with AC$>$5) are on average larger and more
luminous than flocculent (hereafter F) galaxies (i.e. with AC$<$4) (Elmegreen
\& Elmegreen, 1982, 1987).  Late-type spirals are mostly flocculent,
irrespective of their  bar-type, whereas in the early-type spiral range the
percentage of  G objects increases from  $\sim$ 40\% to $\sim$80\% as we go
from unbarred to barred systems (Elmegreen \& Elmegreen, 1989). But in  many
respects G and F spirals are very similar: they do not  appear to differ in
star formation rate --- as  traced  by colours, H$\alpha$ and ultraviolet
fluxes, blue and infrared surface  brighnesses (Elmegreen \& Elmegreen, 1986)
--- in neutral hydrogen content  (Romanishin, 1985), in supernova rate (McCall
\& Schmidt, 1986), in  CO surface brightness (Stark, Elmegreen \& Chance,
1987), in radio  and soft X-ray emissions (per unit light) (Giuricin,
Mardirossian \& Mezzetti, 1989). Recently, Elmegreen \& Elmegreen (1990) and
Biviano et al.  (1991) found that AC correlates with the outer galaxy rotation
curve, in the sense that galaxies with steeper curves tend to have flocculent
arm appearance and galaxies with flatter curves tend to be grand design.  This
would indicate a difference in the relative disk masses (see,e.g., the  review
by Whitmore, 1990), with G galaxies having the largest disk-to-halo mass --- in
agreement with the predictions of the standard density wave theory (see, e.g.,
the textbook by Binney \& Tremaine, 1987).

So far the assessment of environmental influence on the arm structure is one of
the worst known observational aspects of  this topic, since seemingly
controversial results have appeared in the literature. It has been claimed that
the occurrence of G objects in unbarred spirals is greater in spiral members of
pairs and groups than  in relatively isolated systems (Elmegreen \& Elmegreen,
1982) and that,  within the family of galaxy groups, it is greater in the
densest groups (Elmegreen \& Elmegreen, 1983, 1987). On the other hand,
comparing three  samples of interacting galaxies and four samples of galaxy
pairs with  three control samples of bright field galaxies, Giuricin et al.
(1989)  pointed out that F galaxies are more common in the samples of
interacting  or binary systems than in (relatively isolated) field objects.
Noting that  several individual interacting systems display disordered optical
morphology and spatial distribution of neutral hydrogen (e.g., Sulentic \& Arp,
1985 and references cited therein), Giuricin et al. (1989) suggested  that
interactions can destroy G arm structures, thus favouring the occurrence  of F
objects in close galaxy systems. (This is not necessarily in contradiction
with previous claims if the authors refer to very strong interactions only.)

On the theoretical side, the role of the environment in arm  morphology is
generally believed to be important. As a matter of fact,  since the work of
Toomre \& Toomre (1972), close encounters between galaxies  have been
frequently proposed as types of perturbation which can initiate  or maintain
global spiral structure (Kormendy \& Norman, 1979; Toomre, 1981; Sorensen,
1985; Sundelius et al., 1987; Pasha \& Poliachenko, 1987).

The presently confused observational situation has prompted us to  search for
observational evidence of environmental effects on arm structures, using a
rigorous assessment of the environmental density; this is lacking in the
above-mentioned relevant studies, which are content with a fairly  vague
characterization of the environment of the galaxy samples used. In \S 2  we
present the galaxy sample that we have chosen and the parameters that  we have
taken  as good indicators of the local density of galaxies  for each galaxy of
the sample considered. In \S 3 we explore whether AC depends on the local
density of galaxies, analyzing  also correlations betwen AC and other factors
which can influence it, such as the galaxy absolute magnitude and the
disk-to-total mass ratio. As a by-product of our study, we also check whether
barred and unbarred objects  differ in their disk-to-total mass ratio.  In \S 4
we summarize our main results.

\section{The data sample}

In order to describe the environmental density of our nearby  universe we have
chosen the volume-limited NBG catalog (Tully, 1988a), since this galaxy
catalog is intended to include essentially all  the nearby galaxies with
systemic velocities of less than 3000 km/s. This corresponds to a distance of
40 Mpc for the Hubble constant  $H_0$=75 km s$^{-1}$ Mpc$^{-1}$, which value is
adopted throughout the present paper. In the NBG catalog the distances of all
non-cluster  galaxies have been essentially estimated on the basis of
velocities, the above assumed value for $H_0$, and the Virgocentric retardation
model described  by Tully \& Shaya (1984), in which the authors assume the
Milky Way to be  retarded by 300 km/s from the universal Hubble flow by the
mass of the  Virgo cluster. The galaxy members of the groups have been given
a distance  consistent with the mean velocity of the group.

We have updated the Hubble morphological types and the bar-types (SA=unbarred,
SB=barred, SAB=transition-type) of the NBG spiral galaxies by consulting the
RC3 catalog of galaxies (de Vaucouleurs et al., 1992). 511 NBG spiral
galaxies (of which 236 lie at a distance of D$<$20 Mpc) have known arm classes
tabulated in the catalog of Elmegreen \& Elmegreen (1987).

In order to evaluate the local density of galaxies for each of   our 511
spirals, we have to take into account the incompleteness  of the NBG catalog at
large distances. According to Tully's (1988b)  evaluation, the smooth curve
which describes the observed increase in  incompleteness with distance D (in
Mpc) obeys the expression  \be F=\exp[0.041(\mu-28.5)^{2.78}] \ee where
$\mu=5\log D+25$  is the distance modulus and $F=1$ if $\mu<28.5$. The
incompleteness factor $F$ expresses the number of galaxies brighter than
$M_B=-16$ that should  have been cataloged for each object that is listed in
the NBG catalog at  a given distance. Following Tully (1988b) and Giuricin et
al. (1993), we have estimated the contribution  of each galaxy brighter than
$M_B=-16$ to the local density at the specific  locations of our 511 NBG
galaxies with known AC by using a gaussian  smoothing function: \be
\rho_i=C\exp[-r_i^2/2(F^{1/3}\sigma)^2] \ee where $r_i$ is the spatial distance
of the $i$-th galaxy  from the specified location  and the normalization
coefficient is $C=1/(2\pi\sigma^2)^{3/2}=  0.0635/\sigma^3$. For a galaxy at
larger  distance D, the effective smoothing scale length $(F^{1/3}\sigma)$ is
increased in such a way  that the amplitude of the density peak associated with
a galaxy is the same  at all distances D. In such a way the function $\rho_i$
satisfies the normalization  condition  $\int\rho_idV=F$. Then, following
Giuricin et al. (1993), we have defined  the local galaxy density \rs\ of a
given NBG galaxy as the summation over all  contributions of all other galaxies
brighter than $M_B=-16$, excluding the  galaxy itself: \be
\rho_\sigma=\sum_i\rho_i \ee Thus, the local galaxy density \rs\ (that we
express in units of galaxies  Mpc$^{-3}$) is essentially zero (for all
\sg-values) for a galaxy without  companions within a distance of 3\sg\ Mpc.
As discussed in Giuricin et al. (1993),  because of the clustering properties
of galaxies, the choice of different \sg-values  implies a different physical
meaning for the local galaxy density \rs.

In order to calculate the local density \rs\ we need to know the  absolute
magnitudes of the NBG galaxies, those of which with $M_B<-16$ will  be taken as
contributors to \rs. For the NBG galaxies which do not have  \mb\ tabulated in
the catalog (according to their adopted distances and  corrected total blue
apparent magnitudes) we have estimated \mb\ from their  corrected isophotal
diameters $D_{25}$ (relative to the 25 B mag arcsec$^{-2}$  brightness level),
by relying on the following standard luminosity-diameter  relations: \be
M_B=-4.8 \log D_{25} - 13.8 \ee with $D_{25}$ expressed in kpc, for the
elliptical galaxies (Giuricin et al., 1989) and  \be M_B=-5.7 \log D_{25} -
12.4 \ee for the lenticular and spiral galaxies (Girardi et al.,1991).

\section{Analysis And Results}

\subsection{Arm class versus local density and absolute magnitude}

Biviano et al. (1991) have clarified that the AC of a spiral galaxy  depends
mainly on its absolute magnitude \mb\ and its disk-to-total  mass ratio $f$ (as
indicated by the gradients of the outer rotation curve). In the present study
we wish to investigate  whether also  the environmental  density influences the
AC itself. To this end we shall primarily deal  with AC--\rs\ correlations,
taking into account the correlations of AC  with other relevant quantities, in
order to be sure that these  correlations do not induce a spurious AC--\rs\
one.

We analyze the significance of the  correlation between two variables by
computing the linear regression  coefficient \rp\ and the two non-parametric
rank correlation coefficients, Spearman's \ras\ and Kendall's \rk\ (see, e.g.,
Ken\-dall \& Stuart, 1977).  Table 1 lists the three correlation coefficients
and the (one-tailed) percent significance level associated with \rk\ (the one
associated with  \rs\ is generally very similar) for different pairs of
variables and  different subsets of galaxies. The order of the presentation  of
the results in Table 1 follows the discussion reported below.

Subdividing our galaxies  into several morphological  type intervals, we first
analyze the AC--\rs\ correlations. In this case in  which statistics is good we
prefer to restrict ourselves to considering  the subsample of 236 nearby NBG
spirals with D$<$20 Mpc, in order to avoid  possible problems related to the
severe incompleteness of the NBG catalog  at large distances. We find that the
AC--\rs\ correlations turn out to be marginally significant for early types
(for low \sg-values, \sg=0.25 and  0.5 Mpc), in the sense that G objects tend
to inhabit regions of  high local density; for mid types the correlations are
significant for greater \sg-values (\sg=0.5, 1.0, 2.0 Mpc) and in the opposite
sense (F objects tend to stay in denser environments); for late types AC
appears to be unrelated to \rs\ for all \sg-values. In the first lines of Table
1 we report the results just mentioned. The type subdivision used, i.e.,
early-types=S0/a to Sb, mid-types=Sbc to Scd, late-types=Sd to Sm, is the one
which maximizes the positive AC--\rs\ correlations.

Owing to the known strong AC--\mb\ correlation, we divide our nearby early and
mid-type spirals into objects brighter and fainter than \mb=--20 and our nearby
late-type spirals into galaxies brighter and  fainter than \mb=--17.8 (the
adopted limits are close to the medians of the  respective \mb-distributions).
Again no significant AC--\rs\ correlations are  detectable in the late-type
range, whereas  marginal  AC--\rs\ correlations are still observed in  bright
early-type objects, especially when low \sg-values are used. On the other hand,
for the mid-type  range we obtain a strong, positive AC--\rg\ correlation in
the high-luminosity range, together with a strong negative AC--\rg\ correlation
in the low-luminosity range. Fig. 1 presents the AC--\rg\ plots for the
mid-type spirals, subdivided into faint and bright objects. The AC--\rc\
correlation for the mid types shows a similar behaviour, whereas the AC--\rd\
correlation is  significant (and negative) in the low-luminosity range only. We
have also verified that if we omit the 18 galaxies which are members of the
Virgo  cluster (according to the NBG membership assignments), the global
AC--\rs\  correlation for all mid-types vanishes, but the two positive and
negative AC--\rg\ correlations for the high- and low-luminosity subsamples
remain significant. From the physical point of view, we could say that in mid
types the  absence of a significant correlation for \sg=0.25 Mpc indicates that
the density effect on AC is mainly related to the belonging to a galactic
system (like a group or a cluster) rather than to  close companions at a
distance of a few tenths of Mpc; viceversa, in early types the marginal density
effect on AC (for low  \sg-values only) if real could be ascribed to close
encounters between  galaxies.

The extension of this kind of analysis to the whole sample of 128 late-type and
240 mid-type NBG spirals confirms the absence of significant  AC--\rs\
correlations for the former objects and the significance of the  AC--\rs\
correlations (for \sg=0.5, 1.0, 2.0 Mpc) for the latter objects.  On the  other
hand, the study of the entire sample of 136 early-type NBG spirals  does not
confirm the weak AC--\ri\ and AC--\rg\ correlations claimed above. Moreover,
the  analysis of the AC--\rs\ correlations for different bar types shows no
significant bar effect (objects of different bar types behave in the same way).

At this point, after having checked that \mb\ and \rs\ are uncorrelated  for
the NBG nearby  early-, mid- and late-type spirals, we move on to discussing
the important AC--\mb\ correlation. This will also allow us to elucidate the
\mb\ dependence of the AC--\rs\  correlation in mid types. First, it is useful
to check whether this correlation holds for the  whole spiral morphological
sequence. As a matter of fact, one could  speculate that this correlation is a
spurious result of the predominance  of F objects in late-type spirals, which
are on average less  luminous than early types. This view is not correct, since
from the analysis of all nearby NBG  spirals (with D$<$20 Mpc) it turns out
that AC correlates with \mb\ strongly for mid types and marginally  for late
types (in the usual sense), irrespective of their  bar types. Interestingly, in
early types AC appears to be  unrelated to \mb, although closer inspection
reveals a marginal correlation for the SB and SAB early types alone.

Moreover, dividing our mid-type nearby spirals into two ranges  of \rg\ (i.e.
\rg\ greater and smaller than 0.66 gal. Mpc$^{-3}$,  which is the median of the
\rg\ distribution) we find  that the significance of the AC--\mb\ correlation
grows as we move to higher local densities (see Fig. 2). This tendency is also
observed, although in a slightly weaker way, for \sg=1.0 and 2.0 Mpc. Just the
fact that the degree of correlation between AC and  \mb\ depends on the local
density can account for the previously discussed  AC--\rs\ dependence on \mb\
for mid types: when we take bright objects into account, we have more G spirals
in denser environments, and vice versa. No density effect on the AC--\mb\
relation is observed in late types. Some density  effect in the same sense is
detectable for early types (namely,  a weak AC--\mb\ correlation is observed
substantially in systems  located in dense regions; this could justify our
previous finding of a marginal AC--\mb\ correlation for barred early spirals,
which are found in high density regions according to Giuricin et al., 1993).
The analysis of all NBG spirals confirms that for early types AC is generally
uncorrelated with \mb\ and that it correlates slightly  with \mb\ in the
subsets of SB and SAB types; it also confirms the  high degree of correlation
between AC and \mb\ for mid types and leads to a stronger AC--\mb\ dependence
for all late types than for nearby ones.  Besides, the entire NBG sample
provides confirmation of the density effect on the AC--\mb\ relation for mid
types only.

\subsection{Arm class versus fractional disk mass}

Let us now discuss the correlation between AC and the disk-to-total mass  ratio
$f$. This ratio can be computed by means of the logarithmic gradient (LG)  of
the galactic rotation curve. It is defined (e.g., Persic \& Salucci, 1986)   as
\be LG=d\log V(R)/d\log R|_{R_{25}}\, , \ee  where $V(R)$ is the galactic
rotation velocity; LG is measured at the isophotal disk radius $R_{25}$,
deduced from the 25 B mag  arcsec$^{-2}$ isophote. Persic \& Salucci (1990)
related LG to the fractional  disk mass f at the optical radius $R_{25}$ by
means of the following expression: \be f=M_d/M_t=(0.76-LG)/(0.11LG+1.06) \ee
Clearly, $f$ decreases monotonically with increasing LG.

We take the 68 LG-values evaluated by Persic \& Salucci (1993). They have
carefully selected from the literature a sample of 68 spirals, using severe
selection criteria: the chosen galaxies (of types later than Sa, i.e. with no
very prominent bulges) have both exponential light profiles and symmetric
rotation curves extended at least up to 3.2 exponential scale length
(statistically equal to $R_{25}$); besides  the rotation curves
show no evidence of
non-circular motions and  have at least 15 optical measures between 2 and 3
scale length, or are observed with a radio beam size not larger than half a
scale length. This sample of  LG-values can be regarded as the best sample
available so far for a  reliable evaluation of the fractional disk mass, which
we calculate  by means of Equation (7). In the following we denote by $f_1$ the
68 $f$-values evaluated in this manner.

Furthermore, in order to improve the statistics, we consider an  extensive
sample of rough estimates of $f$ derived by Salucci et al. (1993) through a
photometric mass-decomposition method (Salucci, Ashman, Persic, 1991) for 258
spirals suitably selected from the NBG sample. In this method the visible mass
fraction of a galaxy (roughly its fractional disk mass) is obtained by
comparing the visible mass  $M_v=L_B (M/L)_*$ (where $(M/L)_*$ refers to the
stellar component) with the dynamical mass  $M_{dyn}=(V^2(R_{25}) R_{25})/G$,
where $R_{25}$ is the  isophotal radius, $V(R_{25})=W/(2\sin i)$ is the
rotation velocity computed from the 21-cm line width $W$ and $i$ is the
inclination angle of the galactic disk. The stellar mass-to-light ratio
$(M/L)_*$  is obtained from the B--V galaxy colour by means of some published
stellar population synthesis models (Tinsley, 1981; Bruzual, 1983; Jablonbka \&
Arimoto, 1992). In the following we denote by $f_2$ these 258  estimates of
$f$.

Of the above-mentioned samples of 68 and 238 galaxies, 19 and 58  have known AC
and lie at small distances (D$<$20 Mpc). For these galaxies we recognize that
AC does not appear to be appreciably related to $f_1$ or $f_2$ (see Fig. 3 and
Table 1; the weak correlation that seems to be present in fig. 3a is not
statistically significant).
This result is confirmed by extending our analysis to the whole
sample of 24 or 113 galaxies with known AC and known $f_1$ or $f_2$,
irrespective of their distance, morphological  type, bar type, luminosity and
local density. We conclude that the most recent estimates  of $f$ do not
confirm the weak dependence of AC on $f$ suggested by  Elmegreen \& Elmegreen
(1990) and Biviano et al. (1991) on the basis of older data for the rotation
curve gradients. However, our $f_1$ sample is rather poor and our $f_2$ data
are likely to be affected by large uncertanties.

Taking the $f_1$-values for the 35 nearby galaxies (with D$<$20 Mpc) included
in the NBG catalog and the $f_2$-values for the 113 nearby  NBG galaxies (even
if AC is not known), we analyze the $\log f_1$--\rs\ and $\log f_2$--\rs\
correlations. These correlations are never significant for all \sg-values and
for  the usual subdivisions of objects into different morphological  types, bar
types, and luminosity intervals. The analysis of the entire  sample of 47 and
258 NBG galaxies with known $f_1$ and $f_2$ yields the  same conclusion. Thus,
there is no evidence  that the local density influences the fractional disk
mass of a galaxy. In any case, the null results that emerges from the last
correlation  analyses guarantee that the fractional disk mass of a galaxy does
not affect the significance of the density-dependence of the AC--\mb\
correlations discussed above.

\subsection{Arm class and selection effects}

We wonder whether our results could be
substantially affected by observational selection effects. We identify  three
major sources of observational biases: i) the inhomogeneity of the  material
used by Elmegreen \& Elmegreen (1987) for their AC catalog; ii) the inclination
angle $i$ of the galaxy planes (nearly edge-on galaxies   could have AC
determinations of bad quality); iii) the distance of the galaxies (at large
distances the arm classification can suffer from low spatial  resolution).
First, we have checked that the omission of a few galaxies classified  from
high-resolution photographic atlases (rather than from the Palomar  Observatory
Sky Survey) does not affect our main results. Second, we have verified that AC
is unrelated to $i$ for the three subsets of early-type, mid-type, late-type
spirals. Third, within the same three  morphological types, we have seen that
AC does not correlate appreciably with distance, at least for our subsample of
nearby ($D<20$ Mpc) spirals.

\subsection{Bars and fractional disk mass}

A wide variety of theoretical approaches and numerical simulations  have shown
that realistic bar structures, which remain stable over  many orbital periods,
are inevitably formed in unstable galactic disks. On the theoretical side,
basic strategies devised to avoid barlike  instability in a galactic disk rely
on one of the following  requirements: i) galactic rotation curves having an
inner Lindblad resonance; ii) large random motions in the inner parts of the
galactic disks, for instance supplied by a ``hot disk'' component; iii) a
massive halo that contains a mass (at least) comparable to the disk mass (e.g.,
the textbook by Binney \& Tremaine, 1987). An alternative viewpoint proposed by
Lyndell-Bell (1979) envisages  slow formation of bars through the gradual
alignment of stellar eccentric orbits (rather than quick formation as a result
of some  large-scale collective oscillation of disk stars). Moreover, N-body
simulations show that the occurrence of close  encounters between galaxies is a
possible mechanism for  stimulating bar formation in spiral disks (Byrd et al.,
1986; Noguchi, 1987; Gerin, Combes \& Athanassoula, 1990); observational
evidence for this  process is found in early-type spirals only
(Elmegreen, Elmegreen \& Bellin, 1990; Giuricin et al., 1993).

Special attention has been given in the recent literature to the  linear modal
analysis of the global stability of galactic disks. In the context of this
approach, barred objects are expected to  be related to more massive disks than
unbarred galaxies (e.g., the  review by Bertin, 1991). We have seen (see, e.g.,
Persic \& Salucci, 1990) that the disk-to-total mass ratio can be inferred
from the gradient of the outer part of the rotation curve; on the other hand,
the presence of an inner Lindbald resonance depends on the inner part of the
rotation curve, so that the two hypotheses (i) and (iii) are indeed different.
(We recall that the hypothesis (ii) of a hot disk seems to be ruled out by the
observed small velocity dispersion of galactic disks). All this prompts us to
check whether unbarred spirals differ from barred ones in their fractional disk
mass $f$. To this end we consider now the whole samples  of 68 bona fide
$f_1$-values and  258 $f_2$-values mentioned above. The values of \mb\ for the
non-NBG galaxies are calculated by adopting redshift distances with $H_0$=75 km
s$^{-1}$ Mpc$^{-1}$. The $\log f_1$--\mb\ and $\log f_2$--\mb\ plots presented
in Fig. 4,  in which different symbols denote the SA, SAB, and SB galaxies,
illustrate  the known, roughly linear $\log f$--\mb\ correlation (e.g., Persic
\& Salucci, 1990), for 63 and  227 galaxies with known bar type. We have
undertaken an analysis of the correlations $\log f_1$--BAR and $\log f_2$--BAR,
where  the parameter BAR assumes the values 1, 2,  3 for the SA, SAB, and SB
galaxies respectively. The relevant results reported in Table 2 reveal no
appreciable  $\log f_2$--BAR correlation, together with an anticorrelation
between $\log f_1$ and BAR (barred objects would tend to have lower $f$-values
than unbarred ones).

However,  owing to the strong dependence of $f_1$ and $f_2$ on \mb,  it is
necessary in  this case to explore how much of the $f$--BAR correlation (if
any) is not  spurious, namely not induced simply by the strong $f$--\mb\
correlation. To do this, we calculate Kendall's partial correlation coefficient
$r$,  which is a measure of the correlation between two data sets $x$, $y$
independently of their correlation with a third data set $w$ (e.g., Siegel,
1956), where in our case $x$=BAR, $y$=$\log f_1$ or $\log f_2$, $w$=\mb. Since
the sampling distribution of $r$ is unknown, we adopt the bootstrap  method of
resampling (e.g., Efron, 1979; Efron \& Tibshirani, 1985) in  order to compute
its statistical significance, performing 5000  bootstrap resamplings. The
evaluation of $r$ confirms the conclusions drawn above (see Table 2). We
conclude that barred objects certainly do not show greater fractional disk
masses than unbarred ones (at the same luminosity).

\section{Conclusions}

With respect to  earlier studies, our investigation into the factors which
influence the AC of a spiral galaxy evidences a considerable  morphological
type effect on the dependence of AC on \mb. The AC is strongly related to \mb\
(in the sense that G galaxies tend to be more luminous than F objects) in mid
types and, only weakly, in late types; in early types this relation  is very
marginal (it is observed only in barred systems). In view of the lack of
evidence of enhanced star formation in G galaxies (see, e.g., Elmegreen \&
Elmegreen, 1986; Giuricin et al., 1989), this correlation would suggest that
prominent wave modes are more easily generated in bright (large) galaxies. It
is not easy to understand why this should hold substantially in mid types only.

For many years the spectacular examples of some well-known G galaxies with
nearby companions (like M51 or M81) have been regarded as typical cases of G
structures triggered by tidal interactions. Also very recently, a large number
of numerical simulations have been devoted to reproducing G structures by means
of tidal interactions (see, e.g., the extensive survey of computer simulations
by Byrd \& Howard, 1992). However, our statistical study reveals that, amidst
the various mechanisms devised for forming and maintaining spirals --- e.g., i)
modes to feedback cycles and amplification at corotation, ii) edges and grooves
in the density and/or angular momentum distributions, iii) bar-like or oval
potentials, iv) local responses of a galactic disk when forced by a clump, like
a giant molecular cloud, v) tidal perturbations (see, e.g., the review by
Athanassoula, 1990) --- the last one does not seem to be the dominant physical
mechanism at play. As a matter of fact, the influence of the local density on
AC is statistically quite weak. Nevertheless, at variance with earlier
investigations, we find that the local density acts essentially only in
modifying the AC-\mb\ relation  for mid types, making it tighter in denser
environments, whilst no appreciable density effects are detectable in early and
late types. As a consequence, if we select subsamples of bright or faint mid
types, we find positive or negative AC--\rs\ correlations: in this way, for
bright mid types it is indeed true that G spirals are found in denser
environments. Furthermore, previous claims about the greater frequency of F
galaxies in binary/interacting samples than in field galaxy samples (Giuricin
et al., 1989) could be reconciled with the present results if faint galaxies or
late types were overabundant in binary/interacting samples (e.g., the
interacting sample constructed by Keel et al., 1985, shows an excess of late
types).

Using the  most recent estimates of the disk-to-total mass ratio, derived
either from the gradients of good, extended rotation curves or from a
photometric mass decomposition  method, we do not find any significant
influence of this ratio on the AC of spiral galaxies. This finding, which is
seemingly at variance with earlier studies (Elmegreen \& Elmegreen, 1990;
Biviano et al., 1991), is affected by poor statistics ($f_1$-values) and large
uncertanties ($f_2$-values); a larger sample of good  rotation curves is
needed to clarify this issue, which is still to be regarded as an open
question.

We detect no local density effect on the disk-to-total mass ratio.  Our finding
is consistent with the recent results of the  two-dimensional H$\alpha$
observations of  Amram et al. (1992a,b). Obtaining rotation curves from the
analysis of  two-dimensional velocity fields, these authors found that cluster
spirals located in the inner and outer regions of the cluster have  rotation
curves of similar shape. They disclaimed  the view (see, e.g., the review by
Whitmore, 1990), based mainly on slit spectroscopic  observations, that the
rotation curves of spirals near the cluster  center tend to decrease in their
outer parts.

No unambiguous bar effect on the disk-to-total mass ratio emerges from our
study (a different choice of the data sample gives different results); but, in
any case, the results raise some problem for some theoretical scenarios of bar
formation (see, e.g., the review by  Bertin, 1991), which would view barred
galaxies as containing  more fractional disk mass than unbarred systems.

\bigskip
\bigskip

The authors thank Massimo Persic, Paolo Salucci and Elena Pian for enlightening
discussions and for having kindly provided them with the most recent estimates
of disk-to-total mass ratios. The authors are also grateful to Harold G. Corwin
Jr. for having provided them with the ninth tape version of the Third Reference
Catalogue of Bright Galaxies (RC3). This work was partially supported by the
Ministry of University and Scientific and Technological Research (MURST) and by
the Italian Research Council (CNR-GNA).

\newpage
\onecolumn

\section*{Tables}

{\bf Table 1:} Correlation analysis

\begin{center}
\begin{tabular}{r|cccrrrrl}
\hline \hline
N  & $y$ & $x$ & Galaxies & \rp\ \  & \ras\ \ & \rk\ \ & Sign. &  \ \ \ \ \
Notes \\ (1)&(2)&(3)&(4)&(5)&(6)&(7)&(8)\ \ \ &\ \ \ \ \ \ \ (9)\\ \hline
 1 & AC & \ri &  59 &  0.26 &  0.20 &  0.16 & 95.86\% & Early types   \\
 2 & AC & \rg &  59 &  0.16 &  0.20 &  0.15 & 95.70\% & Early types   \\
 3 & AC & \rc &  59 & -0.00 &  0.10 &  0.07 & \nsg    & Early types   \\
 4 & AC & \rd &  59 & -0.02 &  0.05 &  0.03 & \nsg    & Early types   \\
 5 & AC & \ri & 100 & -0.14 & -0.15 & -0.12 & 95.91\% & Mid types   \\
 6 & AC & \rg & 100 & -0.20 & -0.18 & -0.13 & 97.29\% & Mid types   \\
 7 & AC & \rc & 100 & -0.21 & -0.16 & -0.11 & 95.17\% & Mid types   \\
 8 & AC & \rd & 100 & -0.21 & -0.17 & -0.12 & 96.17\% & Mid types   \\
 9 & AC & \rg &  77 &  0.06 &  0.07 &  0.06 & \nsg    & Late types \\
10 & AC & \ri &  36 &  0.31 &  0.16 &  0.12 & \nsg    & Faint early types\\
11 & AC & \ri &  23 &  0.30 &  0.29 &  0.23 & 93.90\% & Bright early types\\
12 & AC & \rg &  59 & -0.39 & -0.46 & -0.35 & \nov    & Faint mid types\\
13 & AC & \rg &  41 &  0.31 &  0.34 &  0.26 & 99.11\% & Bright mid types\\
14 & AC & \mb & 239 & -0.52 & -0.59 & -0.44 & \nov    & All galaxies\\
15 & AC & \mb &  59 & -0.11 & -0.08 & -0.06 & \nsg    & Early types \\
16 & AC & \mb & 100 & -0.52 & -0.54 & -0.40 & \nov    & Mid types   \\
17 & AC & \mb &  77 & -0.06 & -0.14 & -0.11 & 91.49\% & Late types  \\
18 & AC & \mb &  30 &  0.07 &  0.11 &  0.09 & \nsg    & Early, \rg$\leq$0.8\\
19 & AC & \mb &  29 & -0.30 & -0.33 & -0.24 & 96.71\% & Early, \rg$>$0.8\\
20 & AC & \mb &  50 & -0.31 & -0.27 & -0.20 & 97.56\% & Mid, \rg$\leq$0.66\\
21 & AC & \mb &  50 & -0.65 & -0.69 & -0.52 & \nov    & Mid, \rg$>$0.66\\
22 & AC & \lf &  19 &  0.25 &  0.25 &  0.20 & \nsg    & All galaxies \\
23 & AC & \le &  58 &  0.17 &  0.17 &  0.12 & 90.45\% & All galaxies \\
24 & AC & \le &  17 &  0.26 &  0.29 &  0.21 & \nsg    & Early types\\
25 & AC & \le &  41 &  0.06 &  0.09 &  0.03 & \nsg    & Mid types\\
26 & AC & \le &  24 &  0.09 &  0.26 &  0.19 & \nsg    & Faint galaxies\\
27 & AC & \le &  34 &  0.04 &  0.06 &  0.03 & \nsg    & Bright galaxies \\
28 & AC & \le &  29 &  0.07 &  0.12 &  0.08 & \nsg    & \rg$\leq$0.6\\
29 & AC & \le &  29 &  0.22 &  0.19 &  0.13 & \nsg    & \rg$>$0.6\\
30 & AC & $i$ &  59 &  0.01 &  0.04 &  0.02 & \nsg    & Early types\\
31 & AC & $D$ &  59 & -0.09 & -0.06 & -0.06 & \nsg    & Early types\\
32 & AC & $i$ & 100 & -0.07 & -0.09 & -0.07 & \nsg    & Mid types\\
33 & AC & $D$ & 100 & -0.07 & -0.06 & -0.05 & \nsg    & Mid types\\
\hline \hline
\end{tabular}
\end{center}

\newpage

Table 1: (1) progressive number; (2) and (3) parameters considered as dependent
($y$) and independent ($x$) variables in the correlation analysis; (4) number
of
objects; (5) Pearson's linear correlation coefficient \rp; (6) Spearman's rank
correlation coefficient \ras; (7) Kendall's rank correlation coefficient \rk;
(8) one-tailed significance level assigned to \rk; (9) notes: subsample used in
the analysis (all densities are in units of  galaxies per Mpc$^3$).

\bigskip
\bigskip
\bigskip
\bigskip

{\bf Table 2:} Correlation analysis of the fractional mass disk versus bar
structure.

\bigskip

\begin{center}
\begin{tabular}{r|cccrrrr}
\hline \hline
N  & $y$ & $x$ & Galaxies & \rp\ \  & \ras\ \ & \rk/$r$ & Sign. \\
(1)&(2)&(3)&(4)&(5)&(6)&(7)&(8)\ \ \ \\ \hline
 1 & \lf & BAR &  63 & -0.25 & -0.25 & -0.20 & 98.98\% \\
 2 & \lf & BAR &  63 & \multicolumn{2}{c}{const. \mb} & -0.25 & 99.76\% \\
 3 & \le & BAR & 227 &  0.01 &  0.01 &  0.01 & \nsg\\
 4 & \le & BAR & 227 & \multicolumn{2}{c}{const. \mb} & 0.04 & \nsg\\
\hline \hline
\end{tabular}
\end{center}

\bigskip
\bigskip

Table 2: (1) progressive number; (2) and (3) parameters considered as dependent
($y$) and independent ($x$) variables in the correlation analysis (BAR=1 for
SA,
2 for SAB, 3 for SB spirals) ; (4) number of objects; (5) and (6) Pearson's
linear correlation  coefficient \rp and Spearman's rank correlation coefficient
\ras, or third variable held constant in a partial correlation analysis; (7)
Kendall's rank correlation coefficient \rk\ or Kendall's partial correlation
coefficient $r$; (8) one-tailed significance level assigned to \rk\ or $r$.

\newpage
\twocolumn

\newcommand{\apj}{{\it ApJ}}
\newcommand{\apjs}{{\it ApJS}}
\newcommand{\aj}{{\it AJ}}
\newcommand{\mnras}{{\it MNRAS}}
\newcommand{\aea}{{\it A\&A}}
\newcommand{\aeas}{{\it A\&AS}}

\section*{Figure captions}

\bigskip

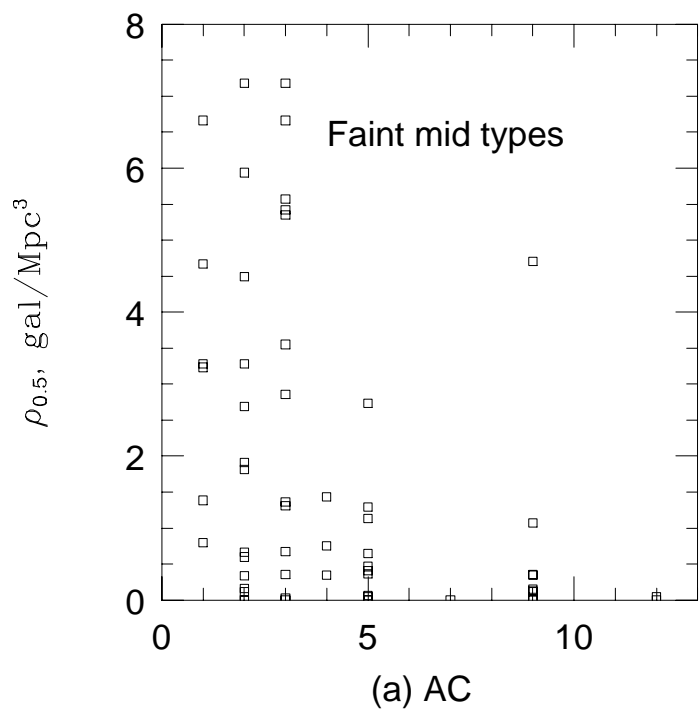
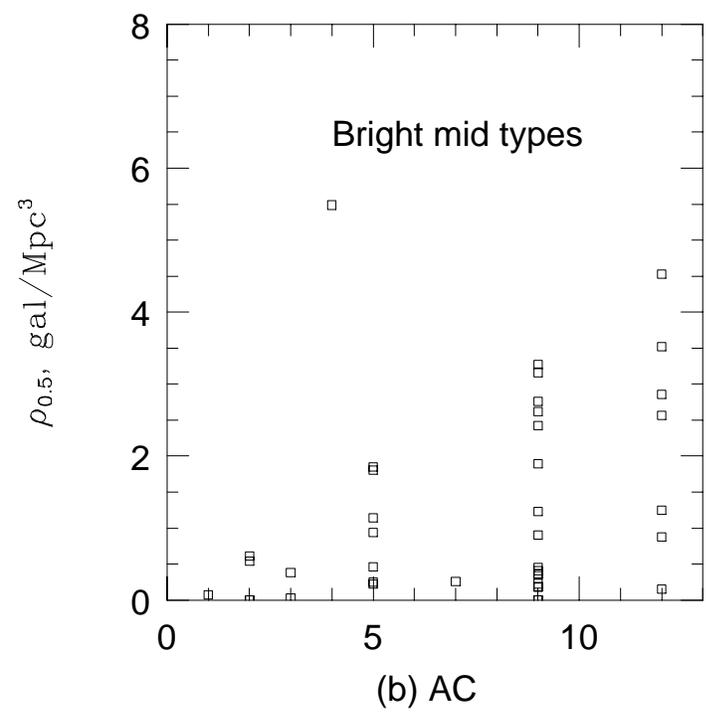
{\bf Figure 1:} \rg\ versus AC plot for mid-type (Sbc to Scd) spirals (a)
fainter and (b) brighter than \mb=--16.

\bigskip
\bigskip

{\bf Figure 2:} \mb\ versus AC plot for mid-type (Sbc to Scd) spirals with
density \rg\ (a) lower and (b) greater than 0.66 galaxies Mpc$^{-3}$.

\bigskip
\bigskip

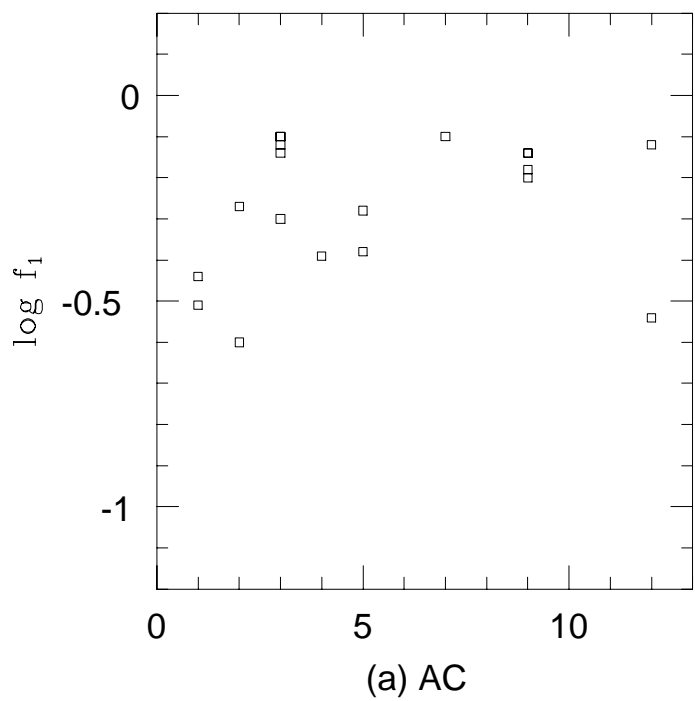
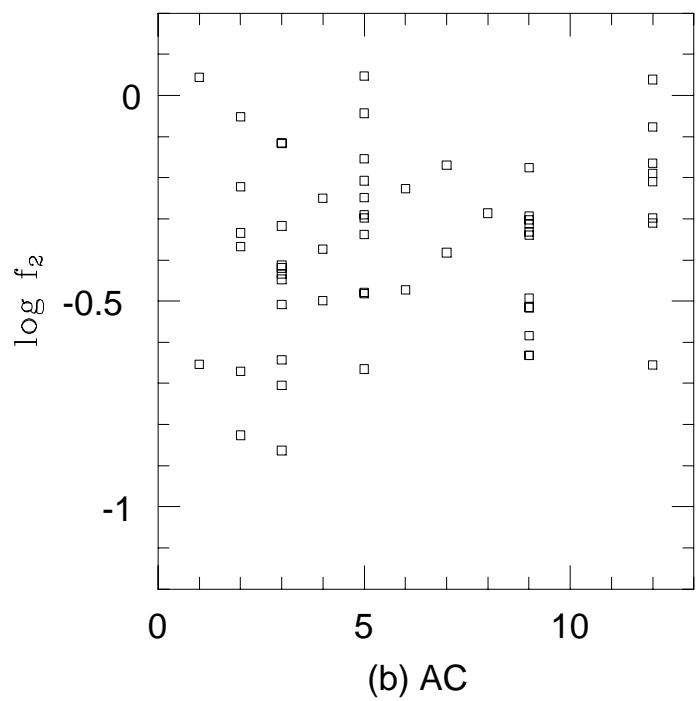
{\bf Figure 3:} (a) $\log f_1$ and (b) $\log f_2$ versus AC plots.

\bigskip
\bigskip

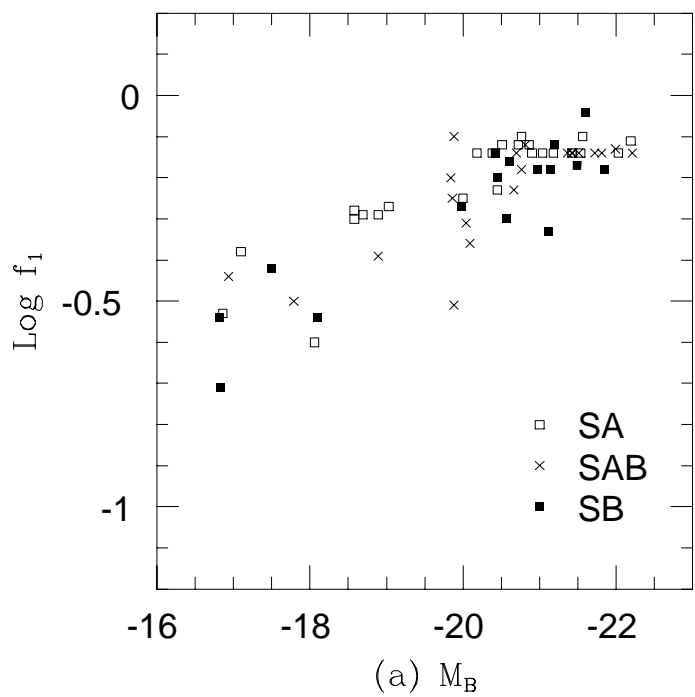
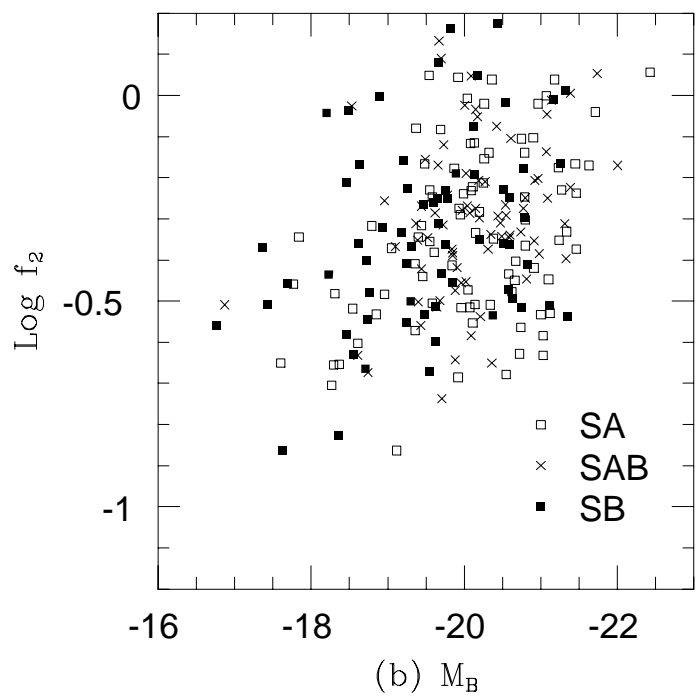
{\bf Figure 4:} (a) $\log f_1$ and (b) $\log f_2$ versus \mb\ plots, with
different symbols for SA, SAB and SB spirals.

\end{document}